\documentstyle[twocolumn,aps,prb]{revtex}
\begin{document}
\draft
\title{Stochastic number projection method in the pairing-force problem}
\author{Roberto Capote\cite{capo}$^1$ and Augusto Gonzalez\cite{augusto}$^{2,3}$}
\address{$^1$Centro de Estudios Aplicados al Desarrollo Nuclear, Calle 30 \# 502, 
 Miramar, La Habana, Cuba\\
$^2$Departamento de Fisica, Universidad de Antioquia, 
AA 1226, Medellin, Colombia\\
$^3$Instituto de Cibernetica, Matematica y Fisica Calle E 309, 
Vedado, Habana 4, Cuba}
\date{\today }
\maketitle

\begin{abstract}
A new stochastic number projection method is proposed.
The component of the BCS wave function corresponding to the right number
of particles is obtained by means of a Metropolis algorithm in which the
weight functions are constructed from the single-particle occupation
probability. Either standard BCS or Lipkin-Nogami probability distributions 
can be used, thus the method is applicable for any pairing strength. 
The accuracy of the method is tested in the computation of pairing energies 
of model and real systems.
\end{abstract}


The microscopic model of superconductivity introduced in 1957 by Bardeen,
Cooper, and Schrieffer (BCS)\cite{BCS} has had astonishing success in
correlating and explaining the properties of simple superconductors in terms
of a few experimental parameters. For its conceptual and computational 
simplicity, the BCS method has
been widely used as the first step in nuclear structure calculations
involving pairing forces\cite{BCS nucleos}. The theory is quite satisfactory
when the number of valence nucleons is large and the pairing interaction is
strong (as compared with the level spacing). But in a nucleus with a
relatively small number of valence nucleons, or with a large spacing 
between levels, the
BCS method fails. As it is well known, the method has two inherent drawbacks:

1) The BCS wave function is not an eigenstate of the number operator. The
energy obtained from this wave function is , therefore, biased with an
inaccuracy caused by the number fluctuation.

2) In some cases, there may be a critical value of the pairing force below
which the BCS equations have no non-trivial solution. Exact calculations
show that this behaviour is spurious\cite{RHO}.

Attempts have been made at improving the method. Kerman et al\cite{KLM}
showed that the wave function obtained by projecting out the BCS function to
the sector with the right number of particles is a very good approximation
to the exact wave function. Many works have been devoted to particle-number
conserving approaches, including projection before and after variation\cite
{UNNA,KAMLAH,LNM,PRADHAM,EGIDO,ZHENG,ESSEBAG}, and many others, see for
example Ref. \onlinecite{RING}. However, it is well known that these
approximate methods can lead, in some cases, to significant errors.

The exact solution of this problem is only available for some very simple
systems, such as a single level or a two-level model\cite{RHO,RING}, or the
case of equispaced doubly degenerate single-particle levels\cite{RICHARD}.
Richardson and Sherman\cite{RICH-SHERMAN} have also developed a general
method for determining the exact eigenvalues and eigenstates of the pairing
force Hamiltonian when the pairing strength is constant. In a recent paper,
Cerf\cite{CERF} has proposed the application of a diffusion Monte Carlo
(DMC) technique to the treatment of the pairing force in nuclei. By making
use of DMC, it is possible to compute exactly the ground-state energy of the
system with a general, state-dependent, pairing interaction, at least in
principle. However, the DMC method can only deal with pairing Hamiltonians
in which the interaction matrix elements are strictly positive. This is
always the case for the nuclear pairing Hamiltonian, but other types of
pairing problems, involving Coulomb interacting particles for example, can
not be solved in this way.

In this paper, we present a novel stochastic algorithm, based on the
Metropolis method\cite{METROP}, for projecting out the component of the BCS
wave function with the correct number of particles. The method does not
depend on the type of interaction involved, as long as we can assume that
the particular BCS functional form holds for our ground-state wave function.
Moreover, number projection can be done starting either from the BCS or
Lipkin-Nogami\cite{LNM,PRADHAM} calculated occupation probability 
distribution. Therefore, it
can be applied for any strength of the pairing interaction. In what follows,
we first introduce the model Hamiltonian, then discuss our Monte Carlo
projection method. Next, we apply the method to simple
systems, and compare the results with those obtained from other approaches.

We consider the problem of a many-body system described by the Hamiltonian,

\begin{eqnarray}
H &=&\sum_{j>0}^{\Omega {}}\epsilon _j\left( a_j^{\dagger }a_j+
a_{\bar{j}}^{\dagger }a_{\bar{j}}\right) -  \nonumber \\
&&-\sum_{j,,j^{\prime }>0}^{\Omega {}}\left\langle j^{^{\prime }}\ 
\bar{j}^{^{\prime }}\left| V_{pair}\right| j\ \bar{j}\right\rangle 
a_{j^{^{\prime}}}^{\dagger }\ a_{\bar{j}^{^{\prime }}}^{\dagger }\ 
a_{\bar{j}}\ a_j
\label{eq1}
\end{eqnarray}

The $a_j^{\dagger }$ and $a_{\bar{j}}^{\dagger }$ create particles in
time-reversed conjugate single-particle states $|j\rangle $ and $|\bar{j}
\rangle $ with energies $\epsilon _j$. The interaction $V_{pair}$ scatters
only time-reversed pairs of particles from the occupied levels 
$|j,\bar{j}\rangle$ to the empty ones $|j^{\prime },\bar{j}^{\prime }\rangle$. 
The indices 
$j$ and $j^{\prime }$ run from 1 to $\Omega $, where $\Omega $ is the total
number of conjugate orbit pairs. Our purpose is to compute the exact
N-particle ground-state energy of $H$. We construct the ground state of $H$
by projecting BCS-type wave-functions onto the N-particle sector. Let us
define the number of pairs, $N_p=N/2$. The following projected
function is obtained:

\begin{equation}
|\Psi _0^N\rangle =C_N\sum_{j_1,\dots ,j_{N_p}}\left[ \prod_{k=j_1}^{j_{N_p}}
 \frac{v_k}{u_k}\ a_k^{\dagger }\ a_{\bar{k}}^{\dagger }\right] \ 
 |\mbox{Vac}\rangle ,
\label{eq2}
\end{equation}

\noindent
where $C_N=\left( \sum_{j_1,\dots ,j_{N_p}}\frac{v_{j_1}^2\dots 
v_{j_{N_p}}^2}{u_{j_1}^2\dots u_{j_{N_p}}^2}\right) ^{-1/2}$ are the 
normalisation coefficients, 
$|\mbox{Vac}\rangle $ is the vacuum state, $v_j$ is the amplitude for 
finding a pair of particles in time-reversed levels, $|j,\bar{j}\rangle$, 
and $u_j$--the amplitude for the levels being empty. These amplitudes are 
obtained either by BCS\cite{BCS} or Lipkin-Nogami (LN)\cite{LNM,PRADHAM} 
methods. Using the Hamiltonian given by Eq. (\ref{eq1}) and the
wave function (\ref{eq2}), we arrive to the following expression for the
projected ground-state energy,

\begin{equation}
E_{BCS}^N=\sum_{j_1,\dots ,j_{N_p}}W(j_1,\dots ,j_{N_p})~\varepsilon
(j_1,\dots ,j_{N_p}),  \label{eq3}
\end{equation}

\noindent
where the sum runs over all possible combinations of $N_p$ occupied
single-particle double degenerated states $\{j_1,\dots ,j_{N_p}\}$, from a
maximum of $\Omega $ states allowed in the Monte Carlo evaluation. The 
``weights'', $W$, and
``energies'', $\varepsilon (j_1,\dots ,j_{N_p})$, are defined as

\begin{equation}
W(j_1,\dots ,j_{N_p})=\frac{v_{j_1}^2\dots v_{j_{N_p}}^2}{u_{j_1}^2\dots
u_{j_{N_p}}^2}\left( \sum_{j_1^{\prime },\dots ,j_{N_p}^{\prime }}
\frac{v_{j_1^{\prime }}^2\dots v_{j_{N_p}^{\prime }}^2}
{u_{j_1^{\prime }}^2\dots u_{j_{N_p}^{\prime }}^2}\right) ^{-1},  
\label{eq4}
\end{equation}

\begin{eqnarray}
&&\varepsilon (j_1,\dots ,j_{N_p})=\sum_{j\in \{j_1,\dots ,j_{N_p}\}}\left(
2\epsilon _j-\left\langle j\ \bar{j}\left| V_{pair}\right| j\ \bar{j}
\right\rangle \right) -  \nonumber \\
&&-\sum_{j\in \{j_1,\dots ,j_{N_p}\}}\sum_{~j^{\prime }\notin \{j_1,\dots
,j_{Np}\}}\left\langle j^{^{\prime }}\ \bar{j}^{^{\prime }}\left|
V_{pair}\right| j\ \bar{j}\right\rangle \frac{u_jv_{j^{\prime }}}
{u_{j^{\prime }}v_j}
\end{eqnarray}

This summation is impracticable as it involves a huge number of terms, 
$\left (\Omega \atop {N_p} \right )$, for usual model spaces. 
Fortunately, the expression (\ref{eq3})
for the projected energy allows a simple Monte Carlo evaluation, where the
ensembles $\{j_1,\dots ,j_{N_p}\}$ are generated with probability 
$W(j_1,\dots ,j_{N_p})$ by means of a Metropolis algorithm\cite{METROP}.
Other equivalent forms of Eq. (\ref{eq3}), see for example Ref. 
\onlinecite{Soloviev}, are not suited for the Monte Carlo evaluation.

In the Metropolis evaluation, we start from the unperturbed ground state and
perform 10$^5$ thermalization steps. We define a trial Metropolis move as a
random transition of one pair from the occupied levels $|j,\bar{j}\rangle $
to the empty ones $|j^{\prime },\bar{j}^{\prime }\rangle $. The trial move
is accepted or rejected according to the Metropolis rule $W(j_1^{^{\prime
}},\dots ,j_{N_p}^{^{\prime }})/W(j_1,\dots ,j_{N_p})>\gamma ,$ where 
$\gamma $ is a random number uniformed distributed between 0 and 1. An
acceptance ratio $R$ is obtained from the thermalization loop. Using this
ratio a total number of Metropolis steps, 
$N_{Metropolis}=300~N_{decorrelation}$ is used to estimate average values of
ground-state energies. The number of decorrelation steps were taken as: 
$N_{decorrelation}=50/R$.

In the following, we will compare results for several soluble models with
those obtained by BCS, LN\cite{LNM,PRADHAM} and DMC\cite{CERF} methods. We
are interested in the two-body pairing energy, $E_{pair}$, which we define as 
$E_{pair}=E_{(V_{pair}=0)}-E_{(V_{pair}\neq 0)}$, where $E_{(V_{pair}=0)}$
and $E_{(V_{pair}\neq 0)}$ are, respectively, the ground-state energies of
the system without and with pairing interaction.

\subsection{The symmetric two-level model}

In this paragraph, we study an exactly solvable, symmetric, two-level model,
with number of particles, $N$, and level degeneracy, $\Omega $. The pairing
interaction is taken constant and equal to $G$. This model has been first
examined by Hogaasen\cite{Hogaasen}, and its exact solution was studied in
detail by Rho and Rassmussen\cite{RHO} in the case $N=\Omega $. More
recently, the general case $N\neq \Omega $ has been discussed in Ref. 
\onlinecite{ZHENG}. The exact pairing energy $E_{pair}$ is obtained by
introducing two sets of quasispin operators, so that the problem finally
reduces to the diagonalization of a tri-diagonal matrix\cite{RHO,ZHENG}. The
results of the BCS approximation, the LN prescription and the DMC
and Metropolis projection methods (MCP) using $v_j$ and $u_j$ either from BCS
(MCP$_{BCS}$) or LN (MCP$_{LNM}$) calculations are compared
against exact results in Table \ref{Table1} for $\Omega =10$ and $N=4,10$
for several values of the interaction strength, $G$. It is shown that
Metropolis Projection methods give very good agreement with the exact
results (the same quality of agreement as the DMC method). Recall that 
for $N\neq \Omega $ the BCS ansatz always have a nontrivial 
minimum\cite{BISHARI}. For $N=\Omega $, a nontrivial solution is found only 
for $G$ greater than $1/(N-1)$. 

\subsection{Equidistant doubly degenerate levels}

In this case, we deal with a system of equispaced doubly degenerate
single-particle levels and a constant pairing interaction. 
This problem has been solved exactly\cite{RICHARD,RICH-SHERMAN} 
for some model spaces with values of the
interaction strength reproducing typical nuclear pairing energies. 
Results obtained from different methods are shown in Table \ref{Table2} 
for $\Omega =N=8$ and three values of $G$. 
Again MCP is in good agreement with the exact results.

\subsection{The $^{100}$Zr nucleus}

In this section, we apply our method to compute the ground-state pairing
energy of the strongly deformed, neutron-rich nucleus $^{100}$Zr. Recently,
the pairing problem in this nucleus was studied by the DMC method\cite
{capote}. The average field was assumed to be an axially deformed
Woods-Saxon potential\cite{28} with Cassinian ovals shape parametrization
\cite{29}. The universal Woods-Saxon parameters proposed by Dudek et al\cite
{22} were used in the single-particle level calculations, except that
smaller values ($R_0=1.25$ for both particles) of the central potential
radius parameter were employed\cite{24}. The computer code CASSINI\cite
{Garrote} using Cassinian ovals parametrization was applied to obtain
single-particle level energies. In the pairing calculations, we keep all
single-particle levels ($\Omega =20$) obtained in the deformed ground-state
potential in a 10 MeV interval around the Fermi level (5 MeV above and 5 MeV
below). The single-particle energies relative to Fermi level energy are
listed in Table \ref{Table3}. The number of pairs in this system is equal to
the number of considered single-particle levels below the Fermi energy, 
i.e. 9 $($number of particles $N=18)$. The BCS , LN and MCP results are 
compared against benchmark DMC calculations in Table \ref{Table4} for
interaction strength $G=0.255$. Differences between published\cite{capote}
and present BCS results arise from the consideration of the self-energy. 
The accuracy of MCP is, once more, comparable to DMC.

\bigskip

In conclusion, we have proposed a method for projecting out BCS-like
wave functions to the $N$-particle Hilbert space. The method shares with 
the BCS approach its computational simplicity. The use of LN approximate 
projection makes it possible to apply the MCP at any pairing strength,
even below the BCS critical coupling, if it exists. These properties 
make the method very suitable for nuclear structure calculations. We
stress that, unlike DMC, any pairing interaction can be treated in 
our approach.

Besides the nuclear pairing problem, there are other possibilities 
of application of the MCP method . Recently, we have used it in the 
computation 
of the ground-state energy of electron-hole systems in a quantum dot
\cite{GONZALEZ}. The electron-hole attractive Coulomb potential is 
the pairing interaction in this case. MCP is used to improve the BCS
estimation. The BCS approach to electron correlations in molecules
\cite{PIRIS,PIRIS1} is another example with Coulomb matrix elements.
It was shown that this approach fails to reproduce the correlation
energy of small molecules, probably because of the need of an exact 
projection. MCP may be a good alternative. Finally, we shall mention 
as a possibility of application the recent study of superconductivity 
in ultrasmall grains\cite{BRAUN}. 

Some of these problems are currently under investigation.

\acknowledgements 
The authors acknowledge support from the Colombian Institute for Science and
Technology (COLCIENCIAS). R. C. is grateful to the Physics Department of the
Universidad de Antioquia for kind hospitality and support.

\begin{table}
\caption{Pairing energies for selected values of the interaction strength in
the two-level model with $\Omega =10$ double degenerate levels and $N=4,10$
particles.}
\label{Table1}
\begin{tabular}{|c|c|c|c|c|c|c|}
G & Exact & BCS & LNM & DMC & MCP$_{BCS}$ & MCP$_{LNM}$ \\ \hline\hline
\multicolumn{7}{|c|}{$\Omega =10, N=4$} \\ \hline
0.024 & 0.202 & 0.161 & 0.205 & 0.205$(3)$ & 0.203$(3)$ & 0.205$(3)$ \\ \hline
0.138 & 1.495 & 1.269 & 1.507 & 1.498$(3)$ & 1.500($5)$ & 1.500$(5)$ \\ \hline
1.105 & 18.05 & 16.276 & 18.05 & 18.05$(9)$ & 18.05($2)$ & 18.05$(2)$ \\ \hline
\multicolumn{7}{|c|}{$\Omega =10, N=10$} \\ \hline
0.024 & 0.147 & 0.000 & 0.122 & 0.146$(2)$ &  & 0.145$(5)$ \\ \hline
0.060 & 0.36  & 0.000 & 0.311 & 0.373$(9)$ &  & 0.353$(9)$ \\ \hline
0.100 & 0.711 & 0.000 & 0.603 & 0.717$(5)$ &  & 0.696$(7)$ \\ \hline
0.500 & 10.56 & 9.300 & 10.56 & 10.56$(5)$ & 10.56$(5)$ & 10.56$(5)$ \\ \hline
1.105 & 28.40 & 25.64 & 28.40 & 28.40$(7)$ & 28.40$(2)$ & 28.40$(2)$ 
\end{tabular}
\end{table}

\begin{table}
\caption{Pairing energies for selected values of the interaction strength in
the system with $\Omega =8$ double degenerate levels and $N=8$ particles.}
\label{Table2}
\begin{tabular}{|c|c|c|c|c|c|c|}
G & Exact & BCS & LNM & DMC & MCP$_{BCS}$ & MCP$_{LNM}$ \\ \hline\hline
0.7 & 5.309 & 3.89 & 5.245 & 5.27$(3)$ & 5.28$(5)$ & 5.25$(6)$ \\ 
0.9 & 8.018 & 6.18 & 7.977 & 8.06($2)$ & 7.98$(5)$ & 7.97$(4)$ \\ 
1.1 & 11.181 & 8.83 & 11.05 & 11.05$(4)$ & 11.11$(5)$ & 11.09$(9)$ 
\end{tabular}
\end{table}

\begin{table}
\caption{Single-particle energies of $^{100}$Zr in a 10 MeV interval around
the Fermi energy. The deformation parameter is $\varepsilon =0.33$.}
\label{Table3}
\begin{tabular}{|l|l|}
Levels above Fermi energy & Levels below Fermi energy \\ \hline\hline 
4.684 & 0.0000 \\ 
4.559 & -0.254 \\ 
4.275 & -1.303 \\ 
3.566 & -1.520 \\ 
3.300 & -2.224 \\ 
2.908 & -2.826 \\ 
2.235 & -3.914 \\ 
1.928 & -4.125 \\ 
1.497 & -4.605 \\ 
1.323 &  \\ 
0.311 & 
\end{tabular}
\end{table}
\medskip 

\begin{table}[tbp]
\caption{Ground-state pairing energy for the interaction strength $G=0.255$
in the $^{100}$Zr nucleus.}
\label{Table4}
\begin{tabular}{|c|c|c|c|c|c|}
G & DMC & BCS & LNM & MCP$_{BCS}$ & MCP$_{LNM}$ \\ \hline
0.255 & 5.0 & 3.87 & 4.94 & 5.1(1) & 4.98(6)
\end{tabular}
\end{table}

\end{document}